\pdfoutput=1
\documentclass[letterpaper,twocolumn,10pt]{article}
\usepackage{usenix2019,epsfig,endnotes}

\usepackage{comment}
\usepackage{multirow}
\usepackage[hyphens]{url}
\usepackage[hypertexnames=false,bookmarks=false,colorlinks]{hyperref}
\usepackage[hyphenbreaks]{breakurl}
\usepackage{xcolor}
\hypersetup{
  colorlinks,
  linkcolor={red!50!black},
  citecolor={blue!50!black},
  urlcolor={blue!50!black}
}
\usepackage[labelfont=bf,font=small,belowskip=-10pt,aboveskip=0pt]{caption}
\usepackage{amsmath,amsopn,amssymb}
\usepackage{endnotes,microtype,xspace,graphicx,fancyvrb,multirow}
\usepackage{booktabs}
\usepackage{array,underscore,relsize}
\usepackage[T1]{fontenc}
\usepackage{times}
\usepackage{fancyhdr,lastpage}
\usepackage{enumitem}
\usepackage{soul}
\usepackage{float}
\usepackage{balance}
\usepackage{textcomp}
\usepackage{subfig}
\usepackage{graphicx}
\usepackage[compact,small]{titlesec}
\usepackage{titling}

\titlespacing*{\section}{0pt}{0.3\baselineskip}{0.3\baselineskip}
\titlespacing*{\subsection}{0pt}{0.2\baselineskip}{0.2\baselineskip}

\usepackage{fp}
\usepackage{siunitx}

\usepackage{xcolor}

\usepackage{pifont}

\newcommand{\sys}{\mbox{\textsc{P\textsuperscript{2}FaaS}}\xspace}

\newcommand{\XXX}[1]{\textcolor{red}{XXX: #1}}

\newcommand{\cc}[1]{\mbox{\smaller[0.5]\texttt{#1}}}

\fvset{fontsize=\scriptsize,xleftmargin=8pt,numbers=left,numbersep=5pt}

\input{code/fmt}

\def\Snospace~{\S{}}

\newif\ifdraft\drafttrue
\newif\ifnotes\notestrue
\ifdraft\else\notesfalse\fi

\input{glyphtounicode}
\newcolumntype{R}[1]{>{\raggedleft\let\newline\\\arraybackslash\hspace{0pt}}p{#1}}

\newcommand{\squishlist}{
\begin{itemize}[noitemsep,nolistsep]
  \setlength{\itemsep}{-0pt}
}
\newcommand{\squishend}{
  \end{itemize}
}

\usepackage{tikz}

\newcommand*\BC[1]{%
\begin{tikzpicture}[baseline=(C.base)]
\node[draw,circle,fill=black,inner sep=0.2pt](C) {\textcolor{white}{#1}};
\end{tikzpicture}}

\usepackage{xstring}
\newcommand{\PP}[1]{
\vspace{2px}
\noindent{\bf \IfEndWith{#1}{.}{#1}{#1.}}
}

\newcommand{\boxbeg}{
\vspace{2px}
\noindent\begin{tabular}{|l|}\hline
\begin{minipage}{3.2in}
\vspace{2px}
\noindent
}

\newcommand{\boxend}{
\vspace{2px}
\end{minipage}\\ \hline
\end{tabular}
\vspace{-10pt}
}

\begin{document}

\title{\sys: Toward Privacy-Preserving Fuzzing \\as a Service}

\author{
  Fan Sang, Daehee Jang, Ming-Wei Shih, Taesoo Kim \\
    Georgia Institute of Technology
}




\date{}

\maketitle

\begin{abstract}
Global corporations (e.g., Google and Microsoft) have recently introduced a
new model of cloud services, fuzzing-as-a-service (FaaS).
Despite effectively alleviating the cost of fuzzing,
the model comes with privacy concerns.
For example, the end user has to trust both cloud and service providers
who have access to the application to be fuzzed.
Such concerns are due to the platform is under the control of its provider
and the application and the fuzzer are highly coupled.
In this paper, we propose \sys, a new ecosystem that preserves end user's
privacy while providing FaaS in the cloud.
The key idea of \sys is to utilize Intel SGX for preventing
cloud and service providers from learning information about the application.
Our preliminary evaluation shows that \sys imposes 45\% runtime overhead
to the fuzzing compared to the baseline.
In addition, \sys demonstrates that, with recently introduced hardware,
Intel SGX Card, the fuzzing service can be scaled up to multiple servers without
native SGX support.

\end{abstract}

\section{Introduction}
\label{s:intro}


Fuzzing-as-a-service (FaaS) is
an emerging paradigm
to accelerate the adoption of
a popular bug finding technique,
fuzzing.
FaaS services,
such as
Google OSS-Fuzz~\cite{google-oss},
Microsoft Security Risk Detection~\cite{microsoft-srd}, and Fuzzbuzz~\cite{fuzzbuzz},
have been designed
to alleviate technical and operational burdens
of integrating fuzzing
as part of software development:
e.g.,
incorporating a well-known fuzzing driver, such as AFL~\cite{afl},
that provides various mutation strategies
for the software under testing,
and scaling it to a large number of servers
on demand.

However, such benefits come
with serious privacy concerns of the software,
hindering the wider adoption of the fuzzing technique.
First of all, 
in the current FaaS model,
developers should trust
the cloud provider, its underlying infrastructure,
as well as operators;
it means that pre-released software
under active development
has to leave out of the developer's complete control.
Second,
since the fuzzing techniques are continuously
discovering 0day vulnerabilities,
malicious (i.e., compromised or curious~\cite{untrusted-cloud})
cloud providers would face some incentives
to monetize the security bugs
without disclosing them to the software developers.
Third,
such limitations are
the concerns of not just the users (i.e., software developers)
but cloud providers as well:
the cloud providers
should spend more operating costs
to manage the security of the infrastructure and the users' software,
as well as to avoid
the damage of the users' reputation
when leaked.

We propose a new ecosystem
that preserves users' privacy while servicing fuzzing on the cloud,
shortly \sys.
The key idea is to 
utilize a modern, commodity
trusted execution environment (TEE),
called Intel SGX,
as the root of trust.
It not only helps FaaS' users
eliminate the cloud provider from
the trust domain,
but also helps the FaaS providers
focus on taking advantage of
the economy of scaling
without worrying
about users' privacy.
We believe fuzzing
can be a killer application
of SGX for two reasons:
1) fuzzing is a CPU intensive task avoiding
the current memory limitation of SGX,
and 2) when the active working set is small,
SGX provides near-native performance
so that developers do not have to trade
performance---fuzzer's great strength---off for privacy.

This paper attempts to draw a picture of an end-to-end ecosystem
of the SGX-enabled FaaS
that addresses the privacy concerns of end-users
as well as
the scalability and early adoption of the fuzzing service.
\sys provides a toolchain
that assists developers
to package their software for testing in a privacy-preserving manner
(\autoref{ss:ocrash})
securely scaling to multiple servers
as necessary
(\autoref{ss:corpus}),
and helps the cloud operators to adopt
it even on a legacy server
with an extension card, Intel SGX Card
(\autoref{ss:vca}).
We make \sys as an open-source project
to inspire and direct current FaaS providers
to foster the fuzzing techniques
for broader audience
in a convenient and privacy-aware manner.

\PP{Summary} This paper makes the following contributions:
\squishlist
\item We propose an end-to-end ecosystem, \sys,
  that adopts, for the first time, SGX and SGX Card,
  envisioning the first steps toward
  implementing the privacy-preserving FaaS.
\item We propose techniques
  to hide crashing information,
  called oblivious crash,
  and to scale it on a larger number of servers
  without minimal, if not none, performance degradation.
\item We make it open source
  to enlighten the FaaS communities
  to show how to address
  the current privacy concerns
  that developers encounter in adopting FaaS.
\squishend

\section{Background and Related Work}
\label{s:bg}


\PP{Fuzzing}
The idea of fuzzing is providing randomly mutated inputs
to a program that aims for triggering abnormal behaviors
such as crashes representing potential bugs.
This process allows fuzzing to execute automatically and
detect bugs with high accuracy.
To improve the effectiveness of fuzzing, one direction is
to increase the throughput of fuzzing. Approaches for this
direction include scaling up the fuzzing with distributed
machines~\cite{clusterfuzz} and designing specialized system support~\cite{wen-ccs17}.
Another direction is to improve the strategy of the seed selection;
that is, inputs derived from a seed that trigger more execution
paths of the program usually lead to more bug findings.

%
%
%
%
%
%
%

%
%
%

\PP{Fuzzing as a service}
Instead of physically owning and maintaining multiple machines for fuzzing,
end-users now have an option to fuzz their applications on the cloud
via "fuzzing as a service" (FaaS).
The idea of FaaS has service providers to set up the fuzzing
infrastructure, typically involves numbers of machines, on the cloud such
that end users can fuzz their applications with a pay-as-you-go model.
The existing service providers include cloud providers themselves
(e.g., Microsoft and Google) and third parties (e.g., Fuzzbuzz and Fuzzit).
Although FaaS makes fuzzing more accessible and successfully finds thousands
of bugs in real-world applications, privacy concerns arise.

\PP{Intel SGX Applications.}
Communities have proposed several SGX-based solutions for preserving privacy
in existing applications.
The examples include network functions~\cite{shih:snfv, poddar:safebricks},
anonymity network~\cite{kim:sgx-tor2, kim:sgx-tor}, and
machine learning~\cite{ohrimenko:oblivious, hunt:chiron}.
To the best of our knowledge, \sys is the first work that uses SGX
to address the privacy problem with FaaS.

\PP{Intel SGX Card}
Although quickly becoming a general security feature in recent desktop-focused CPUs,
SGX receives relatively small support in server-class CPUs that cloud platforms
typically adopt.
To fill this gap, Intel has recently introduced new hardware, the Intel SGX Card.
Intel SGX Card is a re-configured graphic card that consists of three independent,
SGX-capable CPUs.
Similar to a graphic card, the Intel SGX Card is pluggable to multi-socket
server CPUs (connected via PCIe interfaces), and each CPU can have up to four such cards.
As a result, the cards allow both an SGX- and a none-SGX-capable CPU to have additional SGX
support with a high resource density.
Moreover, opposed to buying new SGX-capable machines, adding the cards to existing ones
is more cost- and space-efficient.
These advantages make the Intel SGX Card an optimal, practical option in response to
the growing demand of SGX support in the cloud~\cite{asylo, openenclave}.


\section{Overview}
\label{s:overview}

%
%
%
%

\subsection{Threat Model}
\label{ss:sys-model}

The setting of \sys involves three parties, including
a cloud provider who offers SGX-capable platforms,
a service provider who sets up fuzzers with enclaves in the cloud
platforms and offers fuzzing as a service,
and an end-user who uses the service to fuzz her application.
Among these parties, \sys assumes the cloud provider is the only
untrusted one---either being compromised or simply because of
the other parties do not fully trust---who aims for obtaining
information about the fuzzer and the application which both are
proprietary.
Having full control over the cloud platforms, the cloud provider
can intercept all the incoming and outgoing network traffic.
Further, the cloud provider can freely analyze an initial program
binary to be running inside an enclave.
However, our model assumes the program binary does not contain
memory corruption vulnerabilities that allow for control-flow
hijacking and memory leaks.
Although SGX ensures the confidentiality of the enclave
during its runtime, the cloud provider may still learn information
about the enclave based on observable behaviors such as crashes.
Our model considers side-channel attacks~\cite{gotzfried:cache, xu:cca, lee:sgx-branch-shadow}
against SGX as out of scope.
However, existing side-channel mitigations~\cite{shih:tsgx, gruss:cloak, varys} are applicable
to \sys.

%
%
%

%
%
%
%
%
%
%
%
%
%
%

\subsection{Goals}
\label{ss:goals}

\noindent \textbf{\BC{1} Privacy-preserving.}
\sys considers two types of privacy:
a fuzzing instance and its runtime behavior.
The fuzzing instance
includes a target program
and a fuzzer that provides mutated inputs.
In addition to prevent the cloud provider
from accessing the program and the fuzzer,
\sys also aims for protecting the inputs,
especially for ones that trigger crashes.
Leakage of such inputs indirectly discloses
the vulnerabilities of the program.
%
%
%
Moreover, \sys aims to prevent the cloud provider
from directly inferring the observable runtime behavior,
more specifically, crashes.
%
The occurrence of crashes
directly affects the reputation of end users
(i.e., indicating the number of bugs in the program).
%

\noindent \textbf{\BC{2} Performance.}
Runtime performance directly contributes
to the quality of fuzzing services
(i.e., number of bugs found).
Poor performance reduces
the quality of service
and weakens the motivation of 
using the service.
Therefore,
na\"ively trading performance for privacy
in the case of fuzzing is not acceptable.
\sys aims to maintain the fuzzing performance
while achieving privacy preservation (\BC{1}).

\noindent \textbf{\BC{3} Deployability.}
\sys aims for being deployable to existing cloud platforms.
One aspect for this requirement is that platforms
should provide SGX support that \sys depends on.
For platforms that already provide native SGX support,
such as Microsoft Azure~\cite{azure} and IBM Cloud~\cite{ibm},
\sys is directly deployable.
For platforms provide SGX support via additional hardware
(e.g, Intel SGX Card), \sys should also accommodate
to such environment.
The other aspect is that \sys should allow for easily
adoption; that is, the service provider requires minimal
effort to scale up the fuzzing service across multiple
machines regardless of the type of SGX support (native or
non-native).

\section{\sys Design}
\label{s:design}

\subsection{Workflow}
\label{ss:sys-overview}

\begin{figure}[t]
    \centering
    \includegraphics[width=\columnwidth]{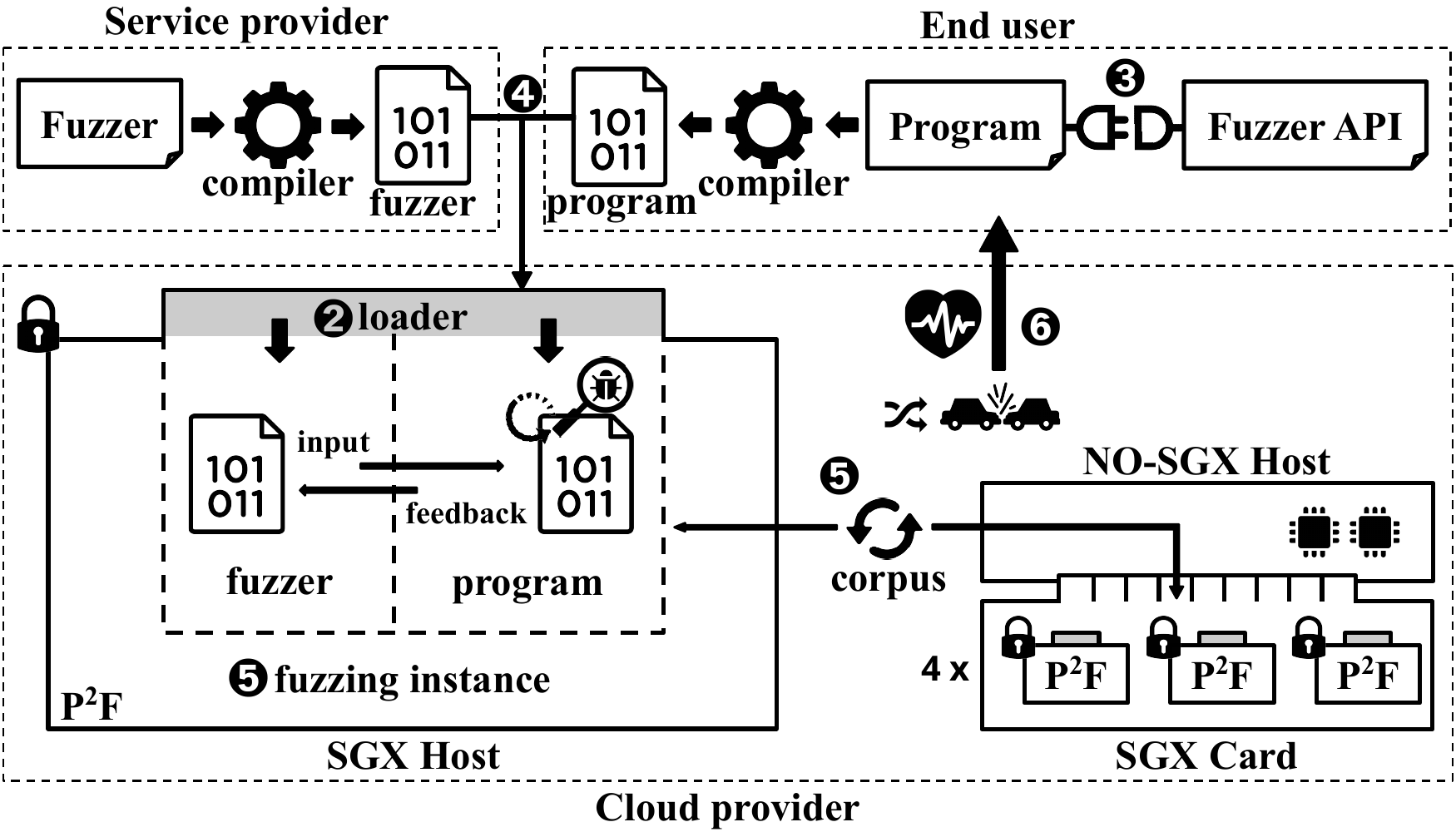}
    \tiny
    \caption{Design overview of \sys.}
    \label{f:design}
\end{figure}

We provide a detailed walk-through of
steps to achieve a complete service cycle with \sys,
as shown in ~\autoref{f:design}.

\noindent \textbf{\BC{1} Fuzzing toolchain.}
The service provider develops a complete
fuzzing tool-chain
in order to provide fuzzing service.
Those tools include,
but not limited to,
a fuzzer and corresponding public APIs,
a compiler that generates binaries
conform to SGX programming practices,
and a loader that loads
both the fuzzer and the target program
into SGX enclave memory.
Although parts of the fuzzing toolchain,
the compiler and loader are not privacy sensitive
but publicly available and verifiable,
serving as trusted anchors
to assist further provisioning of the
fuzzing instance 
into trusted SGX enclave memory.

\noindent \textbf{\BC{2} Service tool provisioning.}
The toolchain loader
developed by the service provider
will be provisioned to
SGX enclaves by cloud providers
before fuzzer APIs are released to the public.
Once provisioned,
the cloud provider will be ready to 
accept fuzzing instance submissions and 
serve end-users with fuzzing service.

\noindent \textbf{\BC{3} Applying fuzzer API.}
The prospective end-user
integrates fuzzer APIs 
released by the service provider into
the target program during development,
and compiles the program
using the toolchain compiler.
Usages of fuzzer APIs vary
across different fuzzers
and is up to the service provider
to specify.

\noindent \textbf{\BC{4} Fuzzing instance provisioning.}
The target program binary
will be submitted to the cloud provider
after being compiled by the end-user.
Meanwhile,
the service provider will be notified
and the corresponding fuzzer binary 
matching the fuzzer API usages of
the target program will also be submitted
to the cloud provider.
The toolchain loader
will then load the fuzzer binary and
the target program binary
into individual SGX local enclaves 
when both are present.

\noindent \textbf{\BC{5} Fuzzing instance execution.}
After both the fuzzer
and target program binaries
are loaded into SGX enclaves,
execution of the fuzzing instance starts.
Necessary service information
is transmitted constantly via
the trusted communication channel between local enclaves.
Specifically,
the fuzzer generates and passes numerous
inputs to the target program
for execution,
and the target program
passes back essential feedback information
to the fuzzer for consumption
(e.g., coverage feedback, crashes).

\noindent \textbf{\BC{6} Fuzzing results.}
Finally,
the fuzzing service periodically
sends fuzzing results (e.g., crashes triggered)
back to the end-user via a 
trusted communication channel
between the cloud provider and the end-user.
The trusted communication channel
can either be established as TLS using
SGX Remote Attestation,
as suggested by Intel~\cite{ra-tls},
or other methods specified by the providers.

\subsection{Oblivious Crash}
\label{ss:ocrash}

The fuzzing instance is naturally protected
since it entirely resides in SGX enclave memory.
However,
program behaviors especially crashes,
require additional protection mechanisms
as the cloud provider is directly aware of such events.
\sys resolves this challenge by
introducing a technique called oblivious crash.
Whose name adopted from oblivious memory,
oblivious crash hides the actual crash
behavior from observers.
That is,
by randomly generating fake crashes
from the target program after fuzzer instrumentation,
the crashing behavior is normalized and 
external observers
cannot distinguish real crashes
from fake ones.

Not only that,
\sys also utilizes fakes crashes as
heartbeat messages to attest 
availability and legitimacy of cloud provider.
On one hand,
fake crash results
with extended time interval indicates
reduced network condition,
and failing to deliver fake crashes periodically
indicates lost of service availability.
On the other hand,
degraded fake crash heartbeat quality
might also indicate potentially malicious
operations are performed,
e.g. the cloud provider pauses the execution
and attempts to distinguish fake crashes.
\sys provides the end user 
with a handful way to sense such situations
upon experiencing different heartbeat behaviors,
and react accordingly.

\subsection{In-Memory Corpus Management}
\label{ss:corpus}
One of the significant performance bottlenecks of SGX application is OCalls.
To minimize OCalls while fuzzing, \sys keeps the corpus and coverage map inside
the enclave as long as the memory can tolerate.
From the initial fuzzer setup, \sys reads in every content of the shared
corpus directory. Afterwards, corpus related operations such as mutating and pruning
are done inside the memory. If the memory is overwhelmed, or the corpus status is
significantly updated, the in-memory corpus is synced with the shared corpus directory.
Ultimately, the shared corpus directory selectively holds the corpus that maximizes
the code coverage.

\subsection{Platform-independent Architecture}
\label{ss:vca}
Similar to traditional fuzzing in parallel~\cite{wen-ccs17},
\sys fuzzing instances can easily run on
multiple SGX-capable nodes in parallel
to scale up execution speed.
However,
as shown in ~\autoref{f:design},
there are two types of platforms
within cloud providers: 
SGX-capable platforms and SGX-incapable ones.
Likely,
instances reside in different cloud providers
consist of a mixture of those two types of platforms.

In order to deploy \sys fuzzing instances
on multiple cloud nodes in parallel
and achieve the performance scalability
without platform restriction,
\sys considers Intel SGX Card as the
most economical and pluggable SGX enabler
for SGX-incapable platforms.
Mutated inputs are shared via the network 
across all \sys fuzzing instances
(shown by \BC{5} in ~\autoref{f:design}),
either hosted on natively SGX-capable platforms,
or hosted on SGX nodes inside Intel SGX Cards.
Therefore, 
\sys achieves platform-independent 
deployability and scalability
by treating each SGX-incapable platforms as
several (3-12) SGX-capable nodes
enabled by Intel SGX Cards.
We further show that
distributing \sys to SGX nodes 
enabled by Intel SGX Card achieves
the same performance scalability 
as by multiple natively SGX-capable instances
in section ~\autoref{s:eval},
proving that Intel SGX Card
is the key to maximize \sys
deployability and scalability.

\section{Implementation}
\label{s:impl}

We implemented the prototype of \sys with roughly 1,000 lines of C code
on top of existing open source projects (SGX-Shield~\cite{seo:sgx-shield}
and libfuzzer~\cite{libfuzzer}).
The prototype includes a fuzzer, a customized toolchain, and an in-enclave
loader.

\PP{Fuzzer}
For simplicity, our fuzzer adopts the model of libfuzzer, which is a part
of the target program, allowing the program to fuzz itself.
We leave the fuzzer decoupling from the program as future work.
Our fuzzer is tailored to dealing with the limitations of SGX (e.g.,
limited memory and no syscall capabilities).
For example, to implement the mutation engine without OCalls,
we use \cc{sgx-rand} for randomness entropy.
Ultimately, the fuzzer exports a call-back API
similar to the fuzz-driver
in libfuzzer.

\PP{Toolchain}
\sys service user should port their code with the provided call-back API;
that is, compiling the program with our toolchain (Clang/LLVM 4.0) with a
new sanitizer option (\cc{-fsanitize=sgxfuzz}) for the LLVM pass.
The compiled program automatically generates both coverage information and
crash reports during the fuzzing.
Our implementation uses 4KB edge coverage map similar to AFL.
To optimize the runtime performance (i.e., minimizing the number of ECalls and OCalls),
we maintain the seed corpus inside the enclave and 
only synchronize it with the file system when needed.


\PP{Oblivious crash}
To implement oblivious crash, we added a fake-crash generator as a part of 
the LLVM pass.
The Fake-crash generator randomly chooses a time between the pre-configured interval.
If a real crash does not occur until the configured time period, 
the instrumented program triggers crash.
The crash is caused by invalid memory access to random address
(using the \cc{sgx-rand} API).
Upon crashes (real or fake ones), the instrumented program sends
out the crash report.
To further hide the crash information, we can encrypt the report.

\PP{Cloud provider}
For \sys cloud provider, we setup an SGX host (Intel i7-6700K) and 
a No-SGX host (Intel Xeon E5-2620) with two Intel SGX Cards.
We pin the fuzzer to each of cores for distributed fuzzing. 
For corpus and coverage synchronization, we use network file system (NFS)
interface.

\section{Evaluation}
\label{s:eval}

In this section,
we evaluate the achievement of privacy preservation,
the usability and performance characteristics of \sys.

\PP{Experiment setup}
We ported three fuzz-drivers: (i) typical toy example for
libfuzzer/AFL (byte-to-byte string match), (ii) arithmetic expression evaluator (30 LoC),
and (iii) C-ARES DNS API from Google fuzzer-test-suite (1,225 LoC). 
Our evaluation numbers are based on an average of ten iterations. 


\subsection{Privacy Preservation}

\begin{figure}
    \begin{center}
    \input{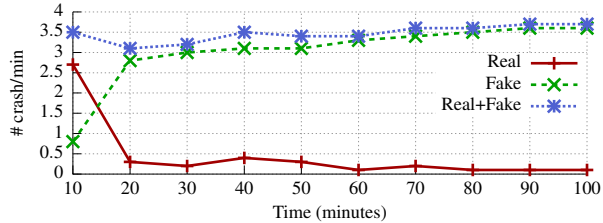}
    \end{center}
    \caption{Oblivious crash behavior (normalized 3 crash/min) while fuzzing the toy fuzz-driver  
    with 50 memory access bugs injected.}
    \label{f:crash}
\end{figure}

While the fuzzing instance is protected
by relying on SGX security guarantees,
\sys target program raises random fake crashes
based on a configurable normalized crash frequency
to obfuscate its runtime behavior (oblivious crash). ~\autoref{f:crash} 
shows the outcome of a normalized 3 crashes per minute.
Specifically,
the frequency of real crashes
starts with a spike and
decreases as time progresses.
Such runtime behavior is normalized
by injecting fake crashes with an inverted frequency distribution.
The resulting observable crash behavior
is represented by \cc{Real+Fake} in ~\autoref{f:crash},
where the crash frequency appears stable over time.
By applying such techniques to normalize crash frequencies,
real crashes are indistinguishable from fake ones,
thus protecting the runtime behavior of the target program.

\subsection{Usability}
One of the major challenges affecting \sys usability is porting the target program into SGX, as SGX does not natively support system calls.
To evaluate \sys with our libfuzzer-based implementation, 
15 OCalls are added in the fuzzer to accommodate SGX environment.
For end users, depending on the target program and fuzzer interface,
number of required OCalls
can significantly differ.
In our evaluation, we selected target programs that do not
invoke system calls to avoid adding OCalls.

For the toy and arithmetic parser examples, the entire programs operate
based on memory and arithmetic operations.
In the case of fuzzer-test-suite example, the parser with
vulnerability (CVE-2016-5180~\cite{cve}) does not require to invoke any system call.
However, depending on the fuzzer interface and program size,
non-trivial porting effort could be involved. We note that these porting efforts can be relaxed with existing techniques~\cite{graphene-sgx}.

\subsection{Performance Characteristics}

\PP{Scaling with Intel SGX Card}
\begin{figure}[t!]
    \begin{center}
    \input{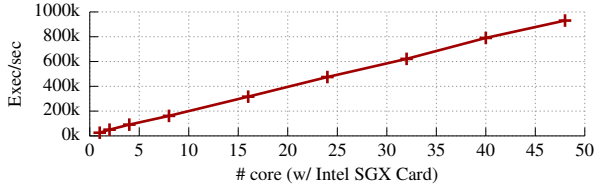}
    \end{center}
    \caption{\sys performance scaling with Intel SGX Card.}
    \label{f:perf}
\end{figure}
To demonstrate the scalability of \sys, we used two Intel SGX Cards to setup  
a 48-core distributed \sys
environment. Fuzzers running on each core share
the corpus and coverage information via network file system (NFS).
Using toy fuzz-driver, we were able to achieve near one million executions per second as shown in~\autoref{f:perf}.

\PP{CPU overhead.}
To measure the CPU overhead of \sys, we show a comparison result with three configurations: 
(i) running the fuzzer without actually using the SGX feature at all (\cc{no-sgx}), 
(ii) running the fuzzer inside enclave with SGX simulation mode (\cc{sgx-sim}), and 
(iii) running the fuzzer inside enclave with SGX hardware mode (\cc{sgx-hw}). 
In particular, we measured the execution time of one million executions. 
Numbers in \autoref{t:speed} are the average of ten iterations.

\begin{table}[t]
    \centering
    \resizebox{\columnwidth}{!} {
        \begin{tabular}{lllc}
\toprule
Fuzz-Driver   & exec/sec (no-sgx)  & exec/sec (sgx-sim) & exec/sec (sgx-hw) \\
\midrule
Toy Example    & 73.63K & 60.22K & 40.32K \\ 
Exp Evaluator  & 72.96K & 60.12K & 40.17K \\ 
C-ARES API     & 72.46K & 59.44K & 39.49K \\ 
\bottomrule
	\multicolumn{4}{l}{\textsc{*}Numbers are based on 1M execution time (10 iteration avg).}
\end{tabular}

    }
    \caption{CPU overhead of \sys}
    \label{t:speed}
\end{table}

The evaluation results show that SGX environment imposes around 45\% performance overhead. 
We suspect this is due to the less optimized cache/TLB implementation.

\PP{Memory overhead.}
Memory usage is another potential cause of performance degradation of \sys under SGX environment. We measure the resident set size (RSS)
with and without \sys using \cc{pmap} to analyze the memory overhead.

Regardless of the target program, the evaluation results give a memory overhead of around \emph{2MB}. The memory overhead is mainly caused by the in-memory 
corpus data structure, coverage bitmap, added fuzzer runtime code, and so forth. 
Such a small memory overhead is ideal for SGX environment as the runtime memory consumption of the fuzzing instance is more likely not to exceed SGX EPC size limitation, thus avoiding the long-worried performance degradation caused by SGX demand paging mechanism.
\section{Discussion and Future Work}
\label{s:discuss}
\PP{Inferring information from fuzzing inputs}
Using SGX allows \sys to protect the target program and
its fuzzing inputs against the cloud provider.
To further prevent the service provider from accessing the
program, \sys puts the program and the fuzzer in separate
enclaves.
However, because the program still relies on the fuzzer
to provide inputs, the service provider may be able
to infer the information about the program from the inputs.
At this point, \sys do not protect programs from
such threats. We leave it to future works.

\PP{Programs with large working set}
Our evaluation test cases used a reasonably small amount of memory, thus
fitting an enclave. 
Currently, size of SGX Enclave Page Cache (EPC) is limited to 128MB.
Instances with larger working set that exceeds the size limitation
will incur the SGX demand paging mechanism,
which is extremely expensive, slowing down the system on
average by 5 times~\cite{vault}.
This is harmful to the service quality of \sys.
We leave the solution to foreseeable future SGX support
for larger EPC size.

\section{Conclusion}
\label{s:conclusion}
Motivated from the rapidly emerging FaaS services,
we propose \sys to provide privacy on top.
Our design demonstrates that SGX is a promising underlying
technology for \sys without harming usability and performance.
As a prototype, we implemented a \sys infrastructure based on
SGX-enabled host and none-SGX host with SGX Card extension.
Our implementation and evaluation demonstrates that \sys is scalable
and deployable in practice.



{\normalsize \bibliographystyle{acm}
\bibliography{p,sslab,conf}}

\end{document}